\documentclass[10pt]{article}

\usepackage{amsmath}
\usepackage{amssymb}
\usepackage{amsthm}
\usepackage{mathabx}

\usepackage{graphicx}

\usepackage{cite}

\usepackage{color} 

\newcommand{\defas}{\buildrel\triangle\over =}
\newcommand{\softmax}[2]{\underset{#1}{{\cal S}{\rm max}} \left({#2}\right)}
\newcommand{\gc}[1]{{\mbox{gain-control}}\left({#1}\right)}
\newcommand{\dt}{distractor mixture}
\newcommand{\ex}{\mathbb{E}}
\newcommand{\nn}{\nonumber}

\newtheorem{theorem}{Theorem}
\newtheorem{conj}[theorem]{Conjecture}
\newtheorem{ext}[theorem]{Extension}

\usepackage[labelfont=bf,labelsep=period,justification=raggedright]{caption}

\makeatletter
\renewcommand{\@biblabel}[1]{\quad#1.}
\makeatother

\date{}

\begin{document}

\begin{flushleft}
{\Large
\textbf{Towards an optimal decision strategy of visual search}
}
\\
Bo Chen$^{}$, 
Pietro Perona$^{\ast}$
\\
Computations and Neural Systems, California Institute of Technology, Pasadena, CA, USA
\\
$\ast$ E-mail: perona@caltech.edu
\end{flushleft}

\section*{Abstract}
Searching for objects amongst clutter is a key ability of visual systems. Speed and accuracy are often crucial: how can the visual system trade off these competing quantities for optimal performance in different tasks? How does the trade-off depend on target appearance and scene complexity? 
We show that the optimal tradeoff strategy may be cast as the solution to a partially observable Markov decision process (POMDP) and computed by a dynamic programming procedure. However, this procedure is computationally intensive when the visual scene becomes too cluttered. Therefore, we also conjecture an optimal strategy that scales to large number of clutters. Our conjecture applies to homogeneous visual search and for a special case of heterogenous search where the signal-to-noise ratio differs across location. Using the conjecture we show that two existing decision mechanisms for analyzing human data, namely diffusion-to-bound and maximum-of-output, are sub-optimal; the optimal strategy instead employs two scaled diffusions. 


\section*{Introduction}

One of the most useful functions of the visual system is searching for things: food, mates, threats. This is a difficult task: the relevant objects, whose appearance may not be entirely known in advance, are often embedded in irrelevant clutter. Furthermore, time is of the essence:  the ability to detect quickly objects of interest is an evolutionary advantage. Speed comes at the cost of making mistakes. Shorter decision times imply collecting less signal, and expose the animal to detection errors. Thus, it is critical that each piece of sensory information is used efficiently to produce a decision in the shortest amount of time while maintaining the probability of errors within an acceptable limit. However, an ideal observer that describes the optimal trade-off between visual search speed (or response time (RT)) and error rate (ER) has not yet been proposed. By ``optimal'' we mean that the observer achieves the lowest expected RT among any alternative model with the same ER. Equivalently an ideal observer is the most accurate among all other models of the same speed. 

Current visual search models fall into two categories. The first category is phenomenological, e.g. diffusion-to-bounds\cite{ratcliff1985theoretical} and competitive accumulators\cite{purcell2012salience}. Such models characterize well the RT versus ER trade-off in humans, but shed little light on optimality. Models of the second category, the ideal observers~\cite{palmer2000psychophysics,verghese01,Chen2011,ma2011behavior}, are optimal but limited only to search tasks with fixed display times. We are interested in ideal observers with unconstrained viewing times,  which 
must solve optimally both the problem of {\it how} to accumulate evidence (for accuracy) and {\it when} to terminate the search (for speed).


One aspect that complicates the ideal observer analysis is that visual search tasks display great variability: the search could be {\it homogeneous}, where target and distractor appearances are known in advance and identical across locations, or {\it heterogeneous}, where some display properties are unknown and/or distinct across locations. Due to this variability, the corresponding ideal observers are often computationally intensive or even intractable. Despite the challenge we make process in three directions. First, we propose a Bayesian framework that describes a lossless evidence accumulation procedure for homogeneous and heterogeneous search problems. Second, we describe a computational solution of ideal observers for visual search that is feasible for small set-sizes (number of search locations), and conjecture an efficient, analytical solution for arbitrary set-sizes. Last, we show that most current visual search models are sub-optimal. To our knowledge, our model is the first that can assess the optimality of existing search models and humans. Moreover, our model characterizes the ER versus RT tradeoff as a function of main parameters that affect search difficulty (e.g. the distinctiveness of the target against the background clutter, the complexity of the image) as well as the degree of uncertainty in these parameters (e.g. varying image complexity from trial to trial). 

Our work differs from previous ideal observer analyses in three aspects. First, we address the problem of finding the optimal time of decision, whereas most ideal observers based on signal detection theory\cite{palmer2000psychophysics,verghese01,Chen2011,ma2011behavior} do not. Second, while ideal observers have been developed for discrimination~\cite{palmer2005effect,gold2001neural}, search is inherently more difficult: input observations in visual search are high-dimensional, and lossless evidence accumulation occurs both locally at each display location and globally across the visual field. In addition, the state space for characterizing the optimal decision strategy is also high dimensional, which makes numerical solvers\cite{drugowitsch2012cost} for the ideal observer impractical and necessitates an analytical solution. Last, we do not model eye-movement, and focus instead on the simpler problem where the observer fixates at all times. This design choice reduces the dimensionality of the optimal decision strategy to a level where an exact solution is tractable. In comparison, models involving eye-movements\cite{chukoskie2013learning,najemnik2008eye} suffer from the curse of dimensionality and thus are rarely optimal under our definition. 


\section*{Results}

\subsection*{Optimal evidence accumulation}
Given sequential observations, an ideal observer is a model that achieves the best ER versus RT trade-off. The ideal observer consists of two components: a process to compute the posterior belief of ``relevant'' (discussed later) variables from the observations, and a process to decide when to stop making new observations. This section is about the first component. We start with existing theories on visual discrimination and homogeneous visual search, and extend them to account for general heterogeneous visual search.

\subsubsection*{Review of visual discrimination}
We first review optimal evidence accumulation for visual discrimination. The stimulus is a single display item, either a distractor (denoted by $C=0$ where $C$ is the stimulus class) or a target ($C=1$). The longer the stimulus is displayed, the more evidence an observer has to assess the target class. In this section we assume the display time is fixed, and the goal is to maximize accuracy given all available evidence. We assume that display items differ in a single attribute $Y$. For simplicity, throughout this paper we assume that the items are tilted bars, and the characteristic attribute is orientation (see Figure.~\ref{fig-setup}(a); the model, however, is completely general and independent of the stimulus design). Target and distractor orientations can take values from two sets $\Theta_1$ and $\Theta_0$, respectively. For example, $\Theta_1=\{30^\circ,10^\circ\}, \Theta_0=\{20^\circ\}$ means that the distractor is $20^\circ$ and the difference between target and distractor orientations (target contrast) is $\pm 10^\circ$. 

The observer receives observation $X(t)$ that grows over time, and we can compute the log likelihood for each stimulus orientation $L_\theta(X(t)) \defas \log P(X(t)|Y=\theta)$. For example, if the input is a gaussian random walk, then $L_\theta(X(t))$ is a scaled version of the input (See Methods~\ref{me:input} or\cite{drugowitsch2012cost}). 
Since here we assume that the optimal length of observation is given, we omit $X(t)$'s dependence on time and write it as $X$. 

Bayesian inference computes the posterior belief of the stimulus class $P(C|X)$. It determines the theoretical error rate: e.g. if $P(C=1|X) = 0.99$, then declaring target-present will produce a $1\%$ error.

The posterior belief can be obtained deterministically from the {\it  log likelihood ratio} $S\defas \log \frac{P(X|C=1)}{P(X|C=0)}$, which is given by~\cite{palmer2005effect}:
\begin{align}
	S_{\mbox{discrim}}(X) &= \softmax{\theta \in \Theta_1}{L_\theta(X)-\log(|\Theta_1|)} - \softmax{\theta \in \Theta_0}{L_\theta(X) - \log(|\Theta_0|)} \label{eq:discrim}
\end{align}
where $|\Theta_1|$ and $|\Theta_0|$ are the number of orientations in the target and the distractor set, respectively;  $\softmax{}{\cdot}$ is the ``softmax'' function, which is the marginalization operation in log probability space: it computes the log probability of a joint event from the log probabilities of its disjoint components:
\begin{align*}
	\softmax{i \in \mathcal{I}}{x} &\defas \log \sum_{i\in\mathcal{I}} \exp(x_i)
\end{align*}

Intuitively, since the target could take one of multiple mutually exclusive orientations, softmax combines evidence from each orientation $L_\theta(X)$ into the log probability for the target and for the distractor, respectively. The two log probabilities are then contrasted to yield the ratio $S_{\mbox{discrim}}(X)$.

\subsubsection*{Review of homogeneous visual search}
In visual search, there are multiple ($M$) items on the display, and at most one of them can be a target. The task is to distinguish between target presence ($C=1$) versus absence ($C=0$). In homogeneous search both the target and distractor orientations $\Theta_1=\{\theta_T\}$ and $\Theta_0=\{\theta_D\}$ are distinct and unique. The log likelihood ratio is~\cite{Chen2011,ma2011behavior}:
\begin{align}
S_{\mbox{homo-search}}(X) &= \softmax{l=1\ldots M}{L_{\theta_T}(X_l) - L_{{\theta_D}}(X_l)} - \log (M) \label{eq:homo-search} 
\end{align}

Since the target could appear at at most one location,  softmax combines the local evidence at disjoint locations into a global log likelihood ratio $S_{\mbox{homo-search}}(X)$ (See~\cite{Chen2011} for derivation). 

\subsubsection*{Search with unknown scene complexity or distractor type}

\begin{figure}[ht!]
\begin{centering}
\includegraphics[width=\linewidth]{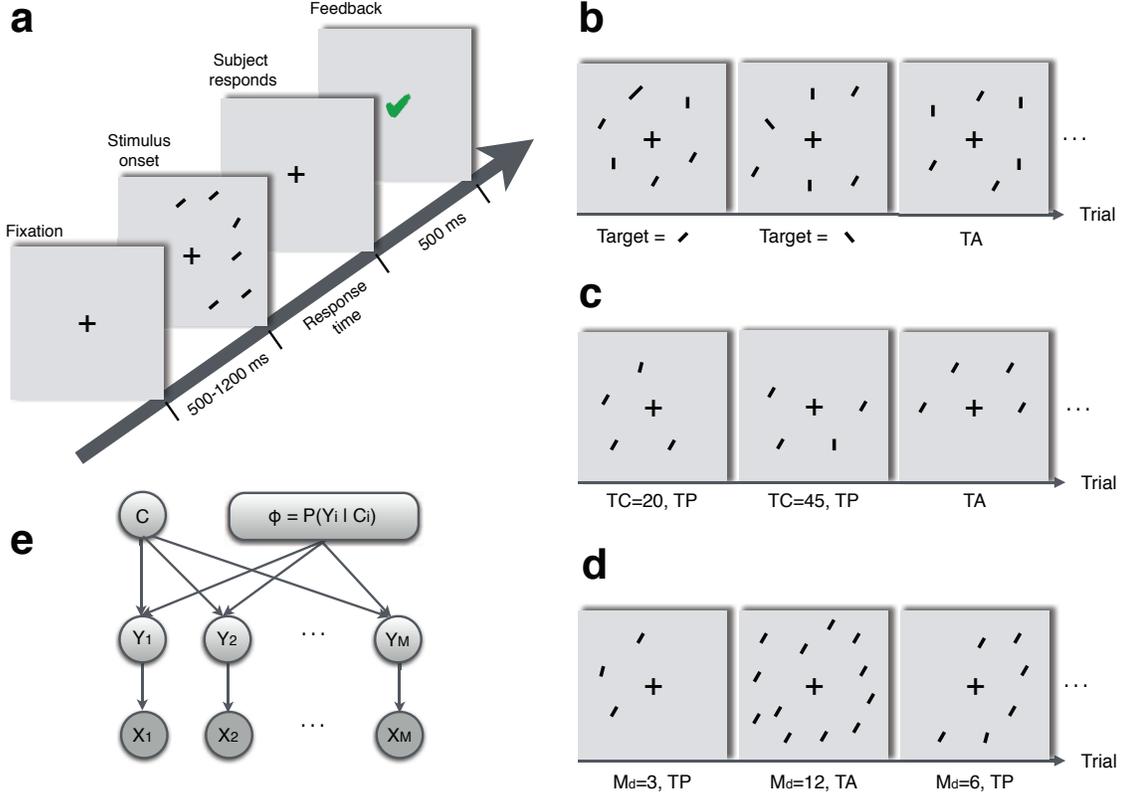}
\caption{{\bf Heterogeneous visual search.} {\bf (a)} Experiment setup. Each trial starts with a fixation screen. Next, a `stimulus' is displayed. The stimulus is an image containing $M$ oriented bars, one of which may be the target. The stimulus disappears as soon as the subject responds by pressing one of two keys, to indicate whether a target is detected or not. Then the screen shows feedback of whether the response is correct, which concludes the trial. Fixation at the center is enforced at all times. {\bf (b-d)} Various heterogeneous search designs. (b) Target and distractor orientations change independently from trial to trial; (c) distractor orientation is tied across all locations, but the target-distractor orientation contrast (TC) varies; (d) mixed number of display items $M_d$. TP: target present; TA: target absent. {\bf (e)} A general graphical model (see details in Equation.~\ref{eq:hetero-search},\ref{eq:hetero-search-target},\ref{eq:hetero-search-distractor} and Methods.~\ref{me:inference}) that explains different designs in (b-d). 
\label{fig-setup}
}
\end{centering}
\end{figure}

In heterogeneous search, some search parameters (e.g. set-size, target contrast) are unknown before stimulus onset. Consequently, an ideal observer must infer these parameters on a trial-by-trial basis. We provide a general framework that encompasses the vast diversity within heterogeneous search, and a Bayesian inference procedure for lossless evidence accumulation.

To unify different heterogenous search tasks we need to define the {\it \dt} $\phi$. The \dt\ encodes the distribution of stimulus orientation $Y_l$ at any non-target location $l$: $\phi_i \defas P(Y_l=\theta_i|C_l=0), \forall l$. Many heterogeneous search tasks may be modeled by considering the effect of the search parameters on the \dt\ (Fig.~\ref{fig-setup}(e), examples below). As each search parameter setting affects the \dt\ differently, $\phi$ is itself a random variable, with distribution $P(\phi), \forall \phi\in\Phi$. We can then specify a heterogeneous search task using the \dt\ distribution. For example:
\begin{itemize}
\item {\it Independent distractors} (Fig.~\ref{fig-setup}(b)). Consider the case where the distractor at each location is sampled independently and uniformly at random from a set of $n$ possible values\cite{ma2011behavior} (e.g. looking for a key on a desk full of different objects). Since at any non-target location, the orientation $Y_l$ could take any of the $n$ values with probability $P(Y_l=\theta|C_l=0) = 1/n, \forall l$, this setting can be described using only one \dt\ $\Phi=\{\phi\}$ where $\phi : \phi_i = \frac{1}{n}, \forall i$. (the corresponding inference~\cite{ma2011behavior} is thus subsumed by our analysis). 
\item {\it Tied distractors} (Fig.~\ref{fig-setup}(c)). Consider the case where the distractors are constrained to be {\it identical} across locations (e.g. looking for a camouflaged prey in a field of tall grass with uniform texture). Here the distractor orientations are deterministic only given the common orientation, which is randomly chosen from $n$ possibilities. More specifically, the \dt\ set uses $n$ \dt s: $\Phi=\{\phi^{(k)}\}_{k=1}^n$, where each \dt\ specifies a deterministic distribution $\phi^{(k)}: \phi^{(k)}_i = \mathbb{I}(k=i)$. 
For example, if $n=3$, then $\Phi=\{[1\ 0\ 0], [0\ 1\ 0], [0\ 0\ 1]\}$. 
\item {\it Unknown scene complexity} (Fig.~\ref{fig-setup}(d)). In the case display items could appear at $M_d$ randomly selected locations out of a maximum of $M$, where $M_d$, the set-size, is sampled uniformly from a discrete set of $n$ values. In this case we introduce ``blanks'' (denoted $\theta_\emptyset$) as a special type of distractor. Any display pattern can be treated as $M_d$ regular items and $M-M_d$ blanks\footnote{In this case the inference procedure is biased: the actual distribution of display locations consists of delta functions at the possible $M_d$ values, while the proposed model implies a mixture of Multinomial distributions, whose means match one-to-one the possible $M_d$ values. Thus the implied distribution is a fuzzier version of the real world, leading to conservative estimates of the log likelihood ratio. In practice, however, blank differentiation is typically instantaneous (i.e. with large diffusion slope), so the consequence of the bias is negligible.}. Thus we need $n$ \dt s, each encodes a distractor probability proportional to the corresponding set-size: $P(Y_l\in\Theta_0|C_l=0)=M_d/M$. For example, say $M=7$, and $M_d$ could be either $1$ or $3$, then $\Phi=\{[1/7,6/7], [3/7,4/7]\}$. 
\end{itemize} 

Given the distribution of \dt s $P(\phi)$, the log likelihood ratio $S_{\mbox{hetero-search}}$ is:

\begin{align}
S_{\mbox{hetero-search}}(X) &= \softmax{l=1\ldots M}{L_{\Theta_1}(X_l) - L_{{\Theta_0}|X}(X_l)} - \log (M) \label{eq:hetero-search} \\
\mbox{where\ \ }L_{\Theta_1}(X_l) &= \softmax{\theta\in\Theta_1}{L_{\theta}(X_l)} - \log(|\Theta_1|) \label{eq:hetero-search-target}\\
L_{{\Theta_0}|X}(X_l) &= -\softmax{\phi\in\Phi}{-\softmax{\theta\in\Theta_0}{L_{\theta}(X_l)+\log\phi_{\theta}}+Q_{\phi}(X)} \label{eq:hetero-search-distractor}
\end{align}

and $Q_{\phi}(X) \defas \log P(\phi|X)$ is the posterior belief of \dt\ $\phi$ given observations from all locations (details later).

The log likelihood ratio expression above is obtained by nesting properly equations from before. At the highest level $S_{\mbox{hetero-search}}$ (Eq.~\ref{eq:hetero-search}) is reminiscent of $S_{\mbox{homo-search}}$ (Eq.~\ref{eq:homo-search}). In addition, since target orientation is unknown, it must be inferred in Equation.~\ref{eq:hetero-search-target} (implemented via a soft-max, as in Equation.~\ref{eq:discrim}). The same applies to distractor orientation, except that the uncertainty is two-fold: both the value and the distribution of distractor orientation are unknown. Hence, two soft-maxes (Eq.~\ref{eq:hetero-search-distractor}), one over the distractor orientation $Y_l$ and the other over the \dt\ $\phi$, are necessary.

The \dt\ $\phi$ can be inferred using gain-control as follows:
\begin{align*}
	Q_{\phi}(X) &= \gc{\log P(\phi) + \sum_{l=1}^M \softmax{\theta\in\Theta_0}{L_{\theta}(X_l) + \log\phi_{\theta}}} \\
	\gc{A_i} &= A_i - \softmax{j}{A_j}
\end{align*}

In conclusion, heterogeneous visual search is challenging because neither the distractor orientation nor its distribution is known in advance. Fortunately, they can be inferred from evidence over the entire visual field, after which the problem reduces to the well-understood, homogeneous visual search. The proposed inference framework incorporates different heterogeneous search modalities such as mixed target type, mixed distractor type and unknown scene complexity. 

\subsection*{Optimal decision strategy}
While the log likelihood makes the best use of existing observations to minimize ER, the ideal observer also requires a stopping strategy in order to trade off RT with ER. The stopping criterion is context-sensitive: an ideal observer should discern whether a situation emphasizes urgency (e.g. competing with other predators for prey) or caution (e.g. avoiding poisonous mushrooms). One common way to capture the importance of error versus time is through a single risk function:

\begin{align}
\mbox{Risk} = ER + C_{\mbox{time}} \mbox{average RT} = \sum_{C=0}^1 \left(P_C(D\neq C)+C_{\mbox{time}}\ex_C[T]\right) \label{eq:risk}
\end{align}
where $D\in\{0, 1\}$ is the observer's decision and $T \in [0\ \ \infty]$ is the response time, respectively, of the sequential test.  $C_{\mbox{time}}$ is the relative cost of time with respect to error\footnote{For simplicity we assume that false positives and false negatives have the same cost, and so do the response times under each class. Different costs can be easily accommodated without affecting the overall analysis.}. The optimal test achieves the lowest risk among all tests. 

\begin{figure}[ht!]
\begin{centering}
\includegraphics[width=1\linewidth]{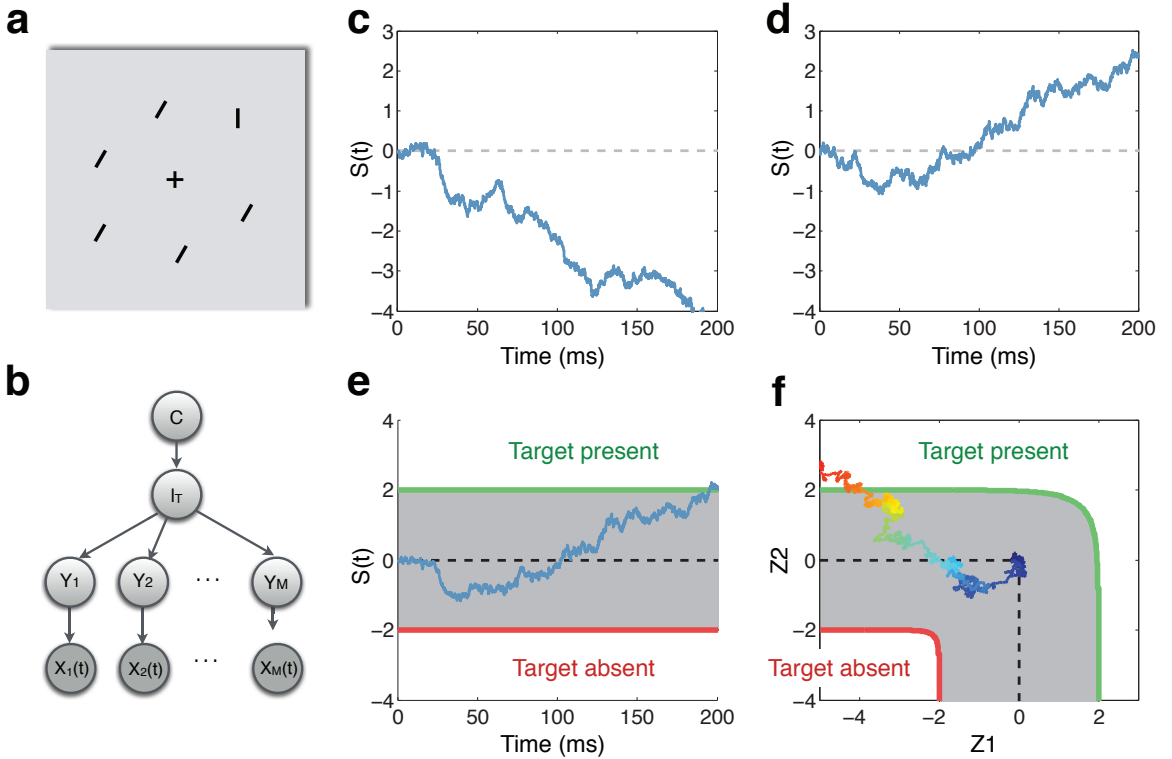}
\vspace{-0.5in}
\caption{{\bf Decision strategies for homogeneous visual search.} {\bf (a)} A homogeneous visual search task: spot the $0^\circ$ target amongst $20^\circ$ distractors. {\bf (b)} A generative model of the problem, simplified from Figure.~\ref{fig-setup}(e). The class variable $C$ (target present or not) controls the location $l_T$ of the target ($l_T=0$ means target absent). $l_T$ in turn determines the item type $Y_{l}$ at each location. Given the item type, the observations $X_{l}$ are independent in space and i.i.d. in time. To perform probabilistic inference, each location computes the local log likelihood ratio $Z_{l}=\log\frac{P(X_{l}|C_{l}=1)}{P(X_{l}|C_{l}=0)}$ over time: {\bf (c)} $Z_{l}$ at a distractor location, {\bf (d)} $Z_{l}$ at the target location. Two decision strategies that make use of the probabilistic interpretation for a two-dimensional visual search problem. SPRT {\bf (e)} thresholds the one-dimensional log likelihood ratio $S(t)$ (Eq.~\ref{eq:homo-search}), whereas the optimal {\bf (f)} uses a decision boundary in the joint space of $\{Z_1, Z_2\}$. Time in (e-f) is color-coded, cooler colors means earlier. 
\label{fig-sample}
}
\end{centering}
\end{figure}

Two components are necessary to describe the optimal test: a state space $\vec{Z}(t)$ over time and a decision strategy that associates each state and time with an action. One common constraint on the state space is that it must be Markov in time: $\vec{Z}(t)$ must be sufficient in summarizing past observations so that given $\vec{Z}(t)$, future observations become independent from the past (see Methods~\ref{me:state}). Once this constraint is satisfied, the problem may be formulated as a partial observation Markov decision process (POMDP)\cite{cassandra1994acting}, and the optimal strategy may be solved exactly using dynamic programming. 

We choose $\vec{Z}(t)$ to be the collection of log likelihood ratios from all locations: 
\begin{align*}
\vec{Z}(t) : Z_l(t) &= L_{\theta_T}(X_l) - L_{{\theta_D}}(X_l),\,\,\,\, l=1\ldots M
\end{align*}

The decision strategy depends on the characteristics of the input $X$. We consider the most common formulation of input as a gaussian random walk at each location (e.g. \cite{thornton2007parallel,drugowitsch2012cost}). The input is parameterized by the {\it drift-rate} $\mu_{C,l}$, which depends on the stimulus class $C$ and the location $l$ (Fig.~\ref{fig-sample}(c-d)). A larger drift-rate difference between the two classes $|\mu_{1,l}-\mu_{0,l}|$ implies a higher signal-to-noise ratio, or equivalently, an easier discrimination problem at location $l$. 

\subsubsection*{Computational solution for low-dimensional problems}
The optimal decision may be computed numerically using dynamic programming~\cite{cassandra1994acting,bellman1956dynamic}. Define $R(\vec{Z}(t), t)$ as the lowest total risk an observer could incur starting from $\vec{Z}(t)$ at time $t$. The optimal risk is equivalent to $R(\vec{0}, 0)$, the total risk from time $0$ onwards with a flat prior. $R(\vec{Z},t)$ is recursively given by:
\begin{align}
R(\vec{Z}(t), t) &= \min \left\{ \begin{array}{cc}
				1-P_0(\vec{Z}(t)) &\quad D=0\text{: declare target absent} \\
				P_0(\vec{Z}(t))  &\quad D=1\text{: declare target present} \\
				C_{\mbox{time}}\delta t + \mathbb{E}_{\vec{Z}({t+\delta t})|\vec{Z}(t)} R(\vec{Z}(t+\delta t),t+\delta t)& \quad D=\emptyset \text{: wait}
				\end{array} \right. \label{eq:dp}
\end{align}
At any time $t$ and any state $\vec{Z(t)}$, the ideal observer picks the action $D\in\{\emptyset,0,1\}$ that yields the lowest risk. If declaring target-absent, the observer makes a false rejection mistake. The false reject probability can be computed from the state $\vec{Z(t)}$ and is denoted $P_0(\vec{Z}(t))$ (see Methods~\ref{me:dp} and Equation~\ref{eq:p1}). If waiting for more evidence, the observer trades off the cost $C_{\mbox{time}}\delta t$ for a new observation of duration $\delta t$, and access to the cumulative risk at $t+\delta t$. 

The optimal decision strategy is defined over a $M+1$ dimensional state-space. The state space is separated by decision boundaries/surfaces into three different decision regions~\cite{sobel1953essentially}. Furthermore, the recurrence equation~\ref{eq:dp} is time invariant. As a result, the optimal decision is constant in time (see~\cite{cassandra1994acting} and Methods~\ref{me:dp}) and the decision surfaces have $M-1$ dimensions. This showcases the difference between the optimal decision strategy for visual search and that for visual discrimination~\cite{palmer2005effect,drugowitsch2012cost}. In discrimination, knowing the target-present probability given past observation is sufficient to compute the likelihood of new observations, hence the state-space is {\it always one-dimensional}. On the contrary, visual search requires the additional knowledge of how the target-present probability breaks down to each location, hence the dimensionality of the state-space is proportional to the set-size.

\subsubsection*{Conjecture for high-dimensional problems} 
\label{sec:conj}
In homogeneous discrimination, the optimal decision strategy is given by the classical Sequential Probability Ratio Test (SPRT)~\cite{wald1945sequential}, which compares the instantaneous log likelihood ratio $S(X(t))$ to a pair of constant thresholds. This strategy has been proven only asymptotically optimal~\cite{dragalin1987asymptotic,schwarz1962asymptotic} for problems that involve multiple hypotheses, such as visual search. 
Nonetheless, in the case of homogeneous visual search, we conjecture that the optimal decision strategy for high-dimensional search is similar to SPRT: in fact it uses two SPRTs defined on scaled log likelihood ratios. 

\begin{conj} (Uniform drift-rates) If all locations share the same drift-rate ($\mu_{1,l}=-\mu_{0,l}=\mu, \forall l$), let $\tau_+$ and $\tau_-$ be the optimal upper and lower thresholds for visual discrimination at location $l$ with a cost of time of $C_{\mbox{time}}$, then the optimal decision surfaces for minimizing the risk function in Equation.~\ref{eq:risk} with the same cost of time $C_{\mbox{time}}$ are:
\begin{align}
	S_+(\vec{Z}(t)) &= \frac{1}{a_+} \softmax{l=1,\ldots,M}{a_+(Z_l(t) -  \log(M))} \geq \tau_+ \label{eq:upper-approx} \\
	S_-(\vec{Z}(t)) &=  \frac{1}{a_-} \softmax{l=1,\ldots,M}{a_-(Z_l(t) - \log(M))} \leq \tau_-   \label{eq:lower-approx} 
\end{align} 
where $a_+$ and $a_-$ are unknown parameters. \label{conj:vs}
\end{conj}

Conjecture \ref{conj:vs} states that the optimal decision strategy is to wait until either $S_+(X(t))\geq\tau_+$ ($D=1$) or $S_-(X(t))\leq \tau_-$ ($D=0$). The thresholds $\tau_+$ and $\tau_-$ are obtained easily by solving a one-dimensional dynamic programming problem~\cite{drugowitsch2012cost}. 
The thresholds are chosen to guarantee asymptotic optimality. Intuitively, when there is only one location ($M=1$), Conjecture.~\ref{conj:vs} reduces to SPRT, which is optimal for visual discrimination. More trickier is the asymptotic case where $M>1$ (multiple locations) but the decision is effectively reduced to concerning only one location $l^*$. This is when all $Z_l(t)$'s at locations $l\neq l^*$ become arbitrarily close to $-\infty$. i.e. when the observer has accumulated a significant amount of information to rule out location $l\neq l^*$ as the target location. In this case:
\begin{align*}
S_{+}(\vec{Z}(t)) &= \frac{1}{a_+} \softmax{l=1,\ldots,M}{a_+(Z_l(t) -  \log(M))} \approx Z_{l^*}(t)-\log(M) \\
S_{-}(\vec{Z}(t)) &\approx Z_{l^*}(t)-\log(M)
\end{align*}
This case happens at location $l^*$ with a probability of $1/M$ when the target is present, and a probability of $1$ when absent. Hence, asymptotically the visual search problem reduces to a visual discrimination problem at location $l^*$ with a log prior ratio of $\log(1/M)$. The best ER vs RT trade-off is achieved when $Z_{l^*}(t)$ is compared against thresholds adjusted by the log prior ratio: $\tau_+ + \log(M)$ and $\tau_-+\log(M)$ (for proof see Methods~\ref{me:thresholds}), which is exactly the asymptotic behavior of Equation.~\ref{eq:upper-approx} and~\ref{eq:lower-approx}.  

\begin{figure}[ht!]
\begin{centering}
\includegraphics[width=0.8\linewidth]{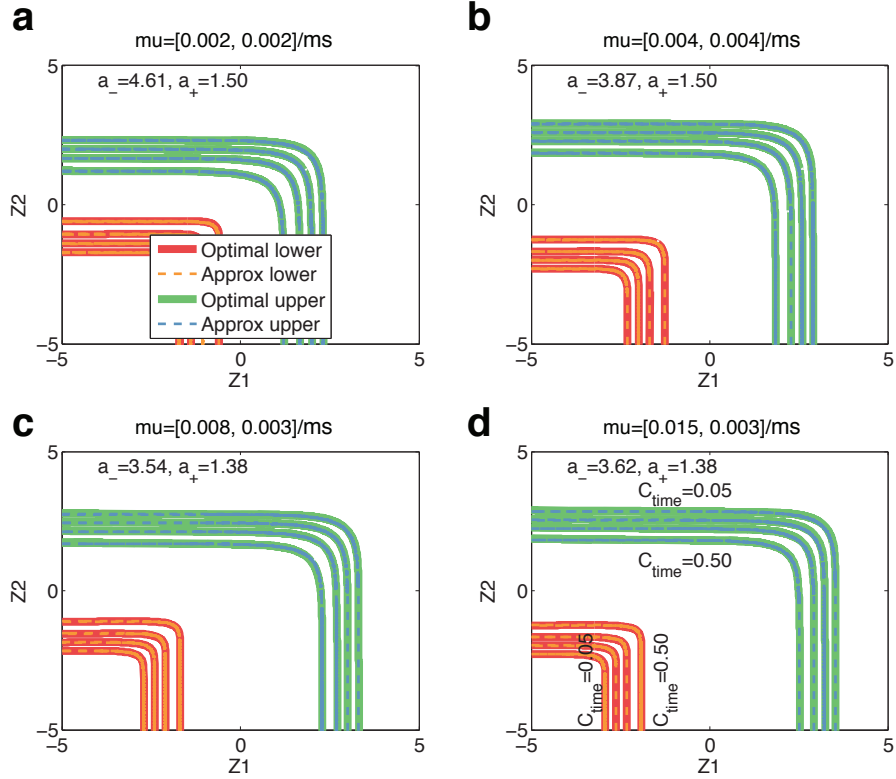}
\caption{{\bf Optimal sequential test for 2D visual search.} {\bf (a-b)} Optimal decision thresholds and approximations for different costs of time $C_{\mbox{time}}\in\{0.5, 0.2, 0.1,0.05\}$ in homogeneous search. Decision boundaries are approximated using Equation.~\ref{eq:upper-approx-diffsnr} and~\ref{eq:lower-approx-diffsnr} with $a_+=1.50$ and $a_-=4.61$. {\bf (c-d)} Optimal decision thresholds and approximations for heterogeneous drift-rate search. Drift-rates are (a-b) $\pm 2/sec$, (c) $\{\pm8, \pm3\}/ms$ and (d) $\{\pm15, \pm3\}/sec$.
\label{fig-optthresh2Ddiffsnr}
}
\end{centering}
\end{figure}

\begin{figure}[ht!]
\begin{centering}
\includegraphics[width=0.8\linewidth]{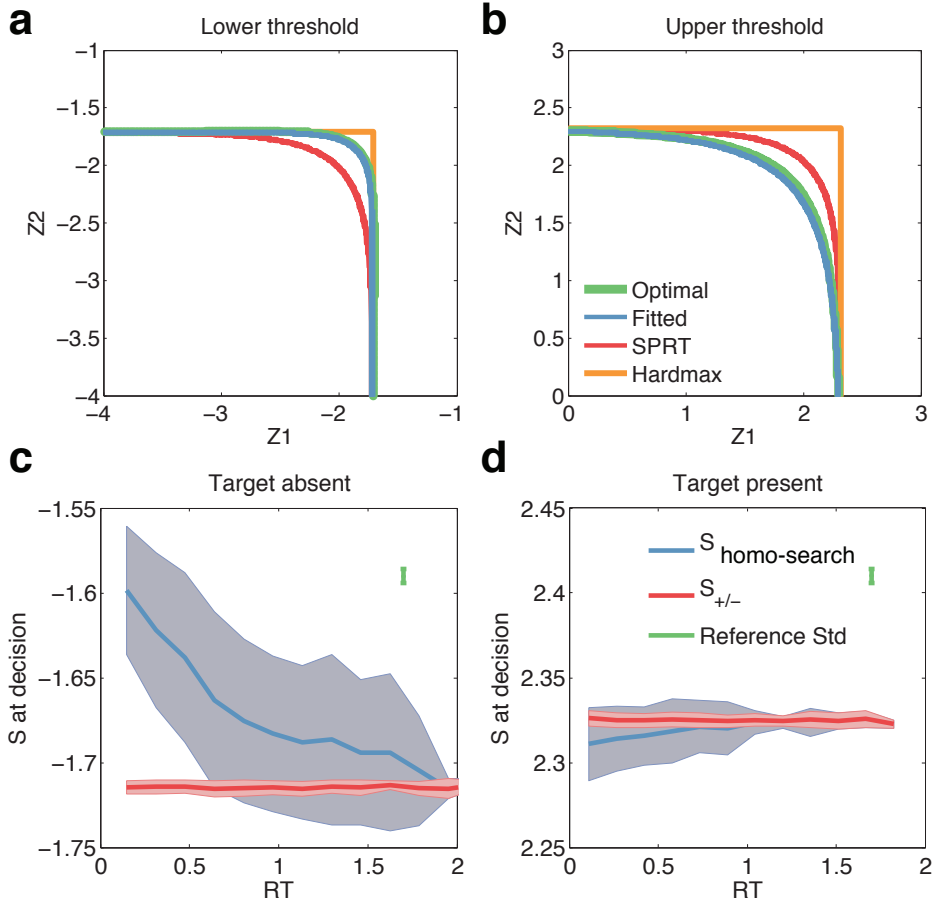}
\caption{{\bf Sequential testing strategies for homogeneous visual search in two-dimensions.} The optimal and various alternative decision strategies are compared in terms of {\bf (a)} the lower and {\bf (b)} the upper threshold in the joint space of $\{Z_1, Z_2\}$. The a-SPRT thresholds are obtained from Equation.~\ref{eq:upper-approx} and~\ref{eq:lower-approx} with  $a_+=1.5$ and $a_-=3.9$; Both SPRT and Hardmax use the optimal threshold for visual discrimination so that asymptotically they are consistent with the optimal strategy. Input to each display location has a drift-rate of $\pm4/sec$. {\bf (c-d)} Each panel shows the log likelihood ratio $S(t)$ distribution at the time of decision under the optimal decision strategy from $1k$ Monte-Carlo simulations. As references, the distribution of $S_-$ when target is absent (c) and of $S_+$ when present (d) are shown. $S_\pm$ is not deterministic because time is discretized in the simulation, which causes the log likelihood ratios to have finite-sized jumps. Standard deviation of the jumps are shown as another reference. Drift-rate of the observation is $\pm2/sec$.
\label{fig-optthresh2D}
}
\end{centering}
\end{figure}

Figure.~\ref{fig-optthresh2Ddiffsnr}(a-b) and Figure.~\ref{fig-optthresh2D} show excellent empirical match between the conjectured thresholds and the optimal thresholds in 2D. 

Our conjecture can be extended to handle a particular heterogeneous search task where the display properties are known, but the drift-rates are different across locations. We refer to this as {\it heterogeneous drift-rate search}, to distinguish it from the much more complex heterogeneous search discussed before. For heterogeneous drift-rate search, Conjecture.~\ref{conj:vs} can be extended by introducing a correction factor for each location:
\begin{ext} (Non-uniform drift-rates) Let $\tau^{(l)}_+$ and $\tau^{(l)}_-$ be the upper and lower thresholds for visual discrimination with a time cost of $C_{\mbox{time}}$ at location $l$, define $c^{(l)}_+ =\tau^{(M)}_+/\tau^{(l)}_+$ and $c^{(l)}_-=\tau^{(M)}_-/\tau^{(l)}_-$,  the optimal decision surface for visual search with the same time cost is: 
\begin{align}
	S_+(\vec{Z}(t)) &= \frac{1}{a_+} \softmax{l=1,\ldots,M}{  c^{(l)}_+a_+(Z_l(t)-\log(M))} \geq \tau^{(M)}_+ \label{eq:upper-approx-diffsnr} \\
	S_-(\vec{Z}(t)) &= \frac{1}{a_-} \softmax{l=1,\ldots,M}{  c^{(l)}_-a_-(Z_l(t)-\log(M))} \geq \tau^{(M)}_-  \label{eq:lower-approx-diffsnr} 
\end{align}
\label{conj:vs_diffsnr}
\end{ext}
The only difference from the uniform drift-rate case (Eq.~\ref{eq:upper-approx-diffsnr}) is that we scale the local diffusions by a location-dependent factor $c^{(l)}_+$. This factor normalizes the diffusion at each location by its relative efficiency with respect to a reference location (arbitrarily chosen to be location $M$). Similar to the intuition that justifies Conjecture.~\ref{conj:vs}, in the asymptotic case where only one location $l^*$ is relevant, $S_{+}(\vec{Z}(t)) \approx \tau^{(1)}_+(Z_{l^*}(t)- \log(M))/\tau^{(l^*)}_+ $. Therefore when the optimal discrimination ER vs RT tradeoff is reached at $Z_{l^*}(t)-\log(M) = \tau^{(l^*)}_+$, the visual search log likelihood ratio $S_+(\vec{Z}(t))$ is at the prescribed threshold $\tau^{(M)}_+$ (Eq.~\ref{eq:upper-approx-diffsnr}). The same argument applies to the lower threshold $\tau^{(l^*)}_-$ and Equation.~\ref{eq:lower-approx-diffsnr}.

Extension.~\ref{conj:vs_diffsnr} only requires solving $M$ one-dimensional dynamic programming problems for $\tau^{(l)}_+$ and $\tau^{(l)}_-$, which is more scalable than the optimal procedure (Eq.~\ref{eq:dp}) that scales exponentially with $M$. Figure.~\ref{fig-optthresh2Ddiffsnr}(c-d) shows that the predicted thresholds from Conjecture.~\ref{conj:vs_diffsnr} match the optimal thresholds from dynamic programming in 2D for a variety of costs of time and drift-rates.

\subsubsection*{Optimality analysis of current search models}
\label{sec:optana}
How are existing visual search models compare against the ideal observer? For fairness we compare only approaches that perform probabilistic inference on the graphical model in Figure.~\ref{fig-sample} (b). These approaches, listed below, differ only in the decision strategy~\cite{palmer2000psychophysics}:

\begin{itemize}
\item {\bf a-SPRT} (Fig.~\ref{fig-sample}(f)): our two-SPRT approach that uses two decision surfaces prescribed in Conjecture.~\ref{conj:vs} and Extension~\ref{conj:vs_diffsnr} to approximate the ideal observer. 

\item {\bf SPRT}~\cite{Chen2011} (Fig.~\ref{fig-sample}(e)): a Bayesian extension of Ward's SPRT~\cite{wald1945sequential} into testing composite hypotheses. SPRT compares the log likelihood ratio of target-present versus target-absent $S(t)$ (Eq.~\ref{eq:homo-search}) against a pair of thresholds. Since the SPRT is subject to the same asymptotic analysis in Conjecture.~\ref{conj:vs}, it uses the same thresholds $\tau_-$ and $\tau_+$ as does the a-SPRT. Essentially, SPRT is a special case of Equation.~\ref{eq:upper-approx} and~\ref{eq:lower-approx} where $a_+=a_-=1$. 

\item {\bf SPRT-opt}: the same as SPRT above except that it optimizes the upper and lower thresholds to minimize the risk function (Eq.~\ref{eq:risk}). Since SPRT-opt may use different thresholds from those in the regular SPRT, it may not be asymptotically optimal. However, this does not prevent SPRT-opt from outperforming the regular SPRT (which is asymptotically optimal). This is because the asymptotic (i.e. long) decisions may only take up a tiny fraction of all the decisions (especially in easy tasks), and SPRT-opt may do better by focusing on the risk for shorter decisions. 

\item {\bf Hardmax}~\cite{Chen2011,najemnik2008eye}: an efficient approximation to SPRT.  Each location decides whether it contains a target $(D_l=1)$ or a distractor $(D_l=0)$ based solely on the local belief $Z_l(t)$. The observer declares target-present when any location reports a target detection, declares target-absent when all locations report a distractor, and waits for more information otherwise. Hardmax is also a special case of Equation.~\ref{eq:upper-approx} and~\ref{eq:lower-approx} where $a_+=a_-=\infty$.
\end{itemize}

{\bf Decision surfaces comparison.}
We want to see how these approaches differ from the optimal in various aspects. First, how different are their decision surfaces? In Figure.~\ref{fig-optthresh2D}(a-b), we compare them on a visual search task with two display locations where it is computationally feasible to solve for the optimal decision boundary using dynamic programming. Since the decision boundaries are constant in time, they can be visualized in the 2-D space of $Z_1$ and $Z_2$ only. Each decision boundary is of the form $\{(Z_1,Z_2)|S(Z_1,Z_2)=\tau\}$, i.e. all pairs of $Z_1$ and $Z_2$ that could make the log likelihood ratio $S$ reach a threshold of $\tau$. 

We observe that both the Hardmax and SPRT {\it differ significantly} from the optimal in terms of the decision surfaces (Fig.~\ref{fig-optthresh2D}(a-b)). SPRT is conservative, because both thresholds bend outwards with respect to the optimal thresholds, which translates to longer decision times for both target-present and target-absent runs. Hardmax, on the other hand, is faster in declaring target-absent but slower in declaring target-present. 

{\bf Can time-varying threshold make SPRT optimal?}
A common practice in modeling decision making in visual discrimination is to employ a time-varying threshold. Can the optimal decision mechanism for visual search also be implemented using SPRT-opt with a {\it time-varying} threshold? 
We reject this hypothesis by computing the $S(t)$ distribution at the time of decision under the optimal test (Fig.~\ref{fig-optthresh2D}(c-d)). If a time-varying threshold exists on $S(t)$ to recover the optimal strategy, the $S(t)$ values should be unique at the time of decision. Instead, we observe a wide spread in the $S(t)$ distribution. 
Therefore, $S(t)$ is not a sufficient statistic to implement the optimal test, and SPRT is sub-optimal in visual search~\cite{wald1945sequential}. 

{\bf Risk comparison.}
The decision surfaces comparison above has one caveat: we consider all places on the decision boundary where decisions {\it could} be taken, ignoring the fact that some places on the boundary are more likely to be reached than others in an actual decision task. E.g. consider Figure.~\ref{fig-optthresh2D}{b} when the search task is easy, the diffusions when the target is present will most likely fall in the region of $\{Z_2>0,Z_1\ll 0\}$ and $\{Z_2\ll 0, Z_1>0\}$, and rarely visit the region of $\{Z_1>0,Z_2>0\}$ where the difference among the strategies is the most noticeable. This reasoning suggests that we should compare these strategies in terms of their actual {\it risk} value. 

\begin{figure}[ht!]
\begin{centering}
\includegraphics[width=0.95\linewidth]{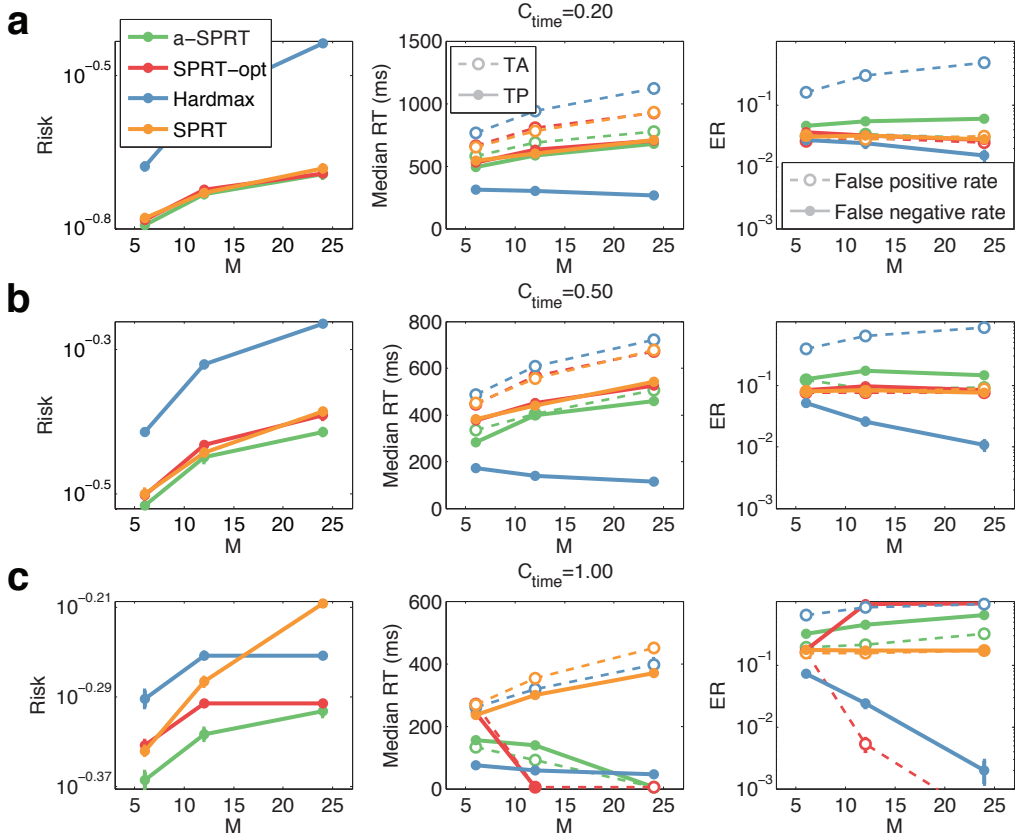}
\caption{{\bf Risk comparison of common decision strategies in homogeneous visual search.} a-SPRT, SPRT-opt, SPRT and Hardmax are compared under different costs of times {\bf (a)} $C_{\mbox{time}}=0.2$, {\bf (b)} $C_{\mbox{time}}=0.5$, and {\bf (c)} $C_{\mbox{time}}=1$ with a drift-rate of $\pm 12/sec$. Hardmax is sub-optimal in all cases. Regular SPRT is sub-optimal in the high cost scenario. SPRT-opt slightly under-performs a-SPRT in terms of the risk. a-SPRT and SPRT-opt are similar in terms of the RT during target-present (TP) and target-absent (TA), as well as the false positive rate and the false negative rate. 
Error bars are one standard error computed from $10k$ runs. 
\label{fig-comp}
}
\end{centering}
\end{figure}

The risks for the strategies in a homogeneous search task are shown in Figure.~\ref{fig-comp}. Hardmax and SPRT are highly sub-optimal. SPRT-opt is almost indistinguishable from a-SPRT in the low time-cost scenario, but becomes sub-optimal when the cost of time becomes very high, i.e. when the decision time is short. Although we have not yet proven that a-SPRT is optimal, it is sufficient to conclude that any model that underperforms it is sub-optimal. 

\begin{figure}[ht!]
\begin{centering}
\includegraphics[width=0.95\linewidth]{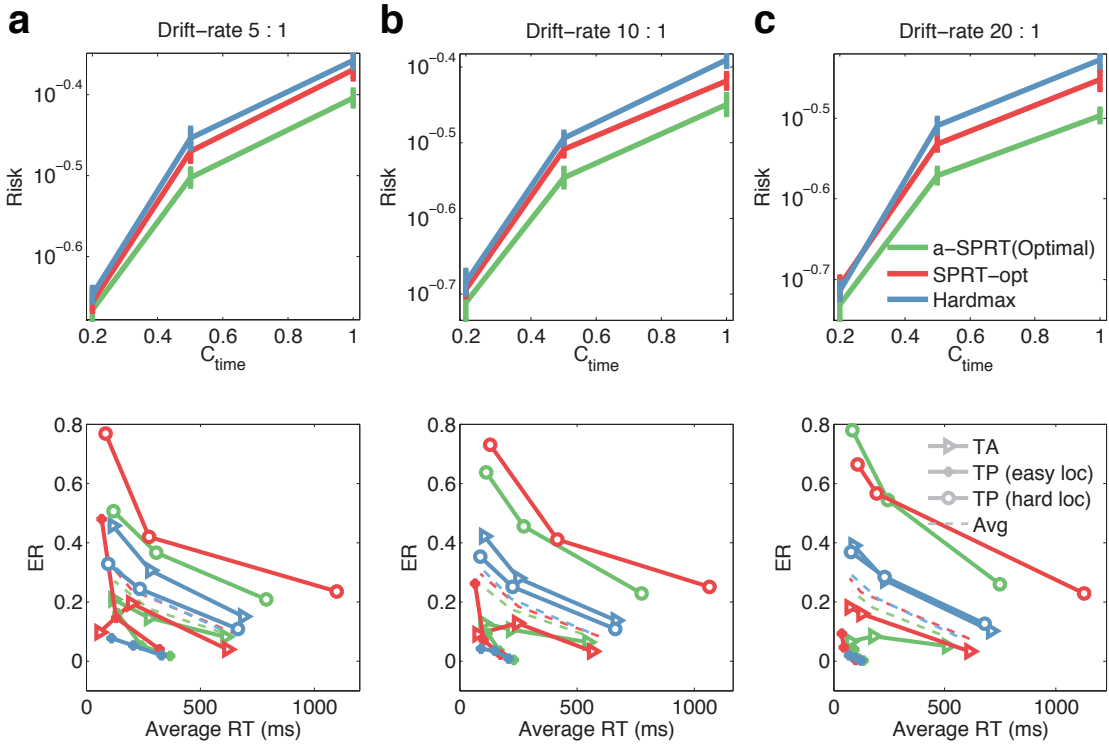}
\caption{{\bf Risk comparison of common decision strategies in heterogeneous drift-rate visual search.} a-SPRT, SPRT-opt and Hardmax are compared under various costs of time. The first row shows the overall risk versus the cost of time. The second row shows the ER vs RT tradeoff under different costs of time (dots) and under three separate conditions (lines): target-absent (TA), target-present (TP) at the location with a larger drift-rate (easy) and target-present at the hard location. Drfit-rates are {\bf (a)} $\{\pm 5,\pm 1\}/sec$,  {\bf (b)} $\{\pm 10,\pm 1\}/sec$ and  {\bf (c)} $\{\pm 20,\pm 1\}/sec$. One standard error in both RT and ER computed from $1k$ runs are shown but too small to be visible. Both SPRT-opt and Hardmax underperform the optimal test. 
\label{fig-comp-diffSNR}
}
\end{centering}
\end{figure}

In heterogeneous drift-rate search (Fig.~\ref{fig-comp-diffSNR}), we see that even with two display locations, both SPRT-opt and Hardmax\footnote{We do not include SPRT because it is not clear how to condense the $M$ asymptotically optimal thresholds, one for each decision surface, into just one for the SPRT. Instead we trust that SPRT-opt, with the ability to optimize the thresholds, should always outperform any SPRT} are suboptimal when the drift-rates differ significantly across locations. The sub-optimality becomes progressively more pronounced as the heterogeneity of drift-rates increases. Behaviorally, when the drift-rate heterogeneity is large, Hardmax achieves near-identical ER vs RT trade-offs at both locations, whereas SPRT-opt and a-SPRT learn to sacrifice the ER at the low drift-rate location for a faster RT overall (Fig.~\ref{fig-comp-diffSNR}{c}). 

In conclusion, decision strategies employed by existing search models are sub-optimal. Hardmax, where one combines local decisions to reach a global decision, is sub-optimal in almost all scenarios. The SPRT-opt, where one executes a one-dimensional SPRT with optimized thresholds, is near-optimal in low cost, homogeneous search scenarios. When the cost of time is high and when the drift-rate is heterogeneous across locations, the SPRT-opt becomes sub-optimal, but remains similar to the optimal strategy in terms of ER and RT.

\section*{Discussion}
We have described the ideal observer model to optimally trade off RT and ER in visual search. The ideal observer relies on lossless evidence accumulation and optimal decision making. We accumulate evidence according to Bayesian inference, where unknown search parameters such as set-size, target contrast and target location are marginalized out. Our model can handle a variety of heterogeneous visual search tasks by modeling the \dt, which is the distribution of distractor orientations. Our model augments the basic search network proposed in~\cite{Chen2011} with a parallel, gain-control circuit that specializes in marginalizing out the \dt s. While the basic search network is common to all tasks, the gain-control circuit is task dependent. This functional separation provides robustness: our model can simply adjust the \dt\ distribution $P(\phi)$ to switch between different search tasks. 

We have conjectured a novel procedure, a-SPRT, to compute the optimal decision policy for high-dimensional homogeneous visual search and heterogeneous drift-rate search. The a-SPRT makes use of two one-dimensional SPRTs with different scaling factors, and with thresholds that are constant in time. In two dimensions, the resultant decision boundary matches closely that of the optimal strategy. The conjecture is preferable over the standard dynamic programming procedure, which does not scale to high (more than three) dimensions. 

Our model enables optimality analysis of humans and phenomenological models of visual search in certain conditions. We compared varies plausible decision strategies and discover that most of them are sub-optimal. While the SPRT with optimized thresholds behaves similarly as the proposed strategy in homogeneous search tasks, it is sub-optimal in search tasks where the signal-to-noise ratio is heterogeneous across locations. 

We highlight several unsolved issues for future work. First, it remains an open question why the optimal decision boundaries for homogeneous search can be described by two scaled-SPRTs. Second, we do not know how the scaling factors $a_+$ and $a_-$ depend on search parameters, and therefore must search numerically for their values to minimize the risk. A better understanding is required to generalize ideal observers of visual search into greater dimensionality and heterogeneity. Third, since both SPRT-opt and a-SPRT can be efficiently implemented using neural hardware~\cite{jazayeri2006optimal,Chen2011}, if Conjecture.~\ref{conj:vs} is proven true, it would imply that the computations required by the ideal observer {\it can} be carried out in cortex. Last, in light of the marked difference between alternative models and the ideal observer in the case of heterogeneous drift-rate search, it would be interesting to test subjects in this case to see which model best captures human behavior, and whether humans {\it are} indeed optimal. 

\bibliography{optimal_visual_search}
\section{Methods}

\subsection{Gaussian input} \label{me:input}
We assume that at each location, the visual system receives gaussian instantaneous observation $\dot{x}(t)$. The cumulative observation at location $l$: $X_l(t) = x(t)$ is a Gaussian random walk:
\begin{align*}
\dot{x}(t) &\sim \mathcal{N}( \mu_\theta \delta t, \delta t)\\
 x(t+\delta t) &= x(t) + \dot{x}(t) \delta t 
\end{align*}
where $\mu_\theta$, the slope of the random walk, depends on the stimulus orientation $\theta$.

At any location $l$, both the log likelihood $\log P(X_l|Y_l=\theta)$ and the log likelihood ratio $\log \frac{P(X_l|Y_l=\Theta_1)}{P(X_l|Y_l=\Theta_0)}$ are linear functions of $x(t)$. First observe that the evidence $X$ at time $t=K\delta t$ is distributed as a gaussian:
\begin{align}
	x(t=K\delta t) &= \sum_{k=1}^{K} \dot{x}(k\delta t) \delta t \sim \mathcal{N}(\mu_\theta t, t) \nonumber \\
	L_\theta(x(t)) &=\log P(x(t)|Y_l=\theta) = \log \mathcal{N}(x(t)|\mu_\theta t, t) \nonumber \\
	&= \mu_\theta x(t) + \frac{\mu_\theta^2t}{2} + \mbox{Const} \label{eq:ll-gaussian-input}
\end{align}
where $\mbox{Const}$ is independent of $\theta$. Similarly,
\begin{align}
	Z_l(t) &= \log \frac{P(x(t)|Y_l=\Theta_1)}{P(x(t)|Y_l=\Theta_0)} \nonumber \\
	&= L_{\Theta_1}(x(t)) - L_{\Theta_0}(x(t)) \nonumber \\
	&= (\mu_{\Theta_1}-\mu_{\Theta_0}) x(t) + \frac{(\mu_{\Theta_0}^2-\mu_{\Theta_1}^2)}{2}t \label{eq:llr-gaussian-input}
\end{align}
Thus both $L_\theta(x(t))$ and $Z_l(t)$ are linear in $x(t)$. In particular we assume the slopes to be symmetrical, i.e. $\mu_{\Theta_1}=-\mu_{\Theta_0}>0$ (otherwise we can adjust the baseline to be $(\mu_{\Theta_0}+\mu_{\Theta_1})t/2$ and make the slopes symmetrical), then $Z_l(t) =  2\mu_{\Theta_1} x(t)$ is directly proportional to $x(t)$. 

\subsection{Bayesian inference for heterogeneous visual search} \label{me:inference}
The target-present likelihood is given by marginalizing out the target location $l_T$, distractor mixture $\phi$, as well as the target and distractor orientations:
\begin{align*}
P(X|C=1) &= \sum_{l_T, \phi} P(X|l_T, \phi, C=1) P(\phi) P(l_T|C=1) \\
&= \sum_{l_T} P(l_T|C=1) \sum_\phi P(\phi) \sum_{\vec{Y} = \{Y_1,\ldots,Y_M\}}P(X|\vec{Y}) P(\vec{Y}|l_T, \phi,C=1) \\
&= \sum_{l_T} P(l_T|C=1) \sum_\phi P(\phi) \sum_{\vec{Y}}\prod_l (P(X_l|Y_l) P(Y_l|l_T, \phi,C=1)) \\
&= \sum_{l_T} P(l_T|C=1) \sum_\phi P(\phi) \prod_l\sum_{Y_l} (P(X_l|Y_l) P(Y_l|l_T, \phi,C=1)) \\
&= \sum_{l_T} P(l_T|C=1) \sum_\phi P(\phi) P(X_{l_T}|C_{l_T}=1) \prod_{l\neq l}P(X_l|\phi,C_l=0) \\
&= \sum_{l_T} P(l_T|C=1) \sum_\phi P(\phi) \frac{P(X_{l_T}|C_{l_T}=1)}{P(X_{l_T}|\phi,C_{l_T}=0)}\prod_i P(X_l|\phi, C_l=0)
\end{align*}
where
\begin{align*}
 P(X_l|C_l=1) &= \sum_{\theta\in\Theta_0}P(X_l|Y_l=\theta)P(\theta|C_l=1) \\
 P(X_l|\phi,C_l=0) &=\sum_{\theta\in\Theta_0}P(X_l|Y_l=\theta) \phi_\theta
\end{align*}

Similarly, the target-absent likelihood is:
\begin{align*}
P(X|C=0) = \sum_\phi P(\phi) \prod_l P(X_l|\phi, C_l=0)
\end{align*}

Define:
\begin{align*}
 P(\phi|X) &= \frac{P(\phi)\prod_lP(X_l|\phi,C_l=0)}{\sum_{\phi'} P(\phi') \prod_l P(X_l|\phi', C_l=0)}
 \end{align*}
 
 Then the log likelihood ratio is:
 \begin{align*}
 \log\frac{P(X|C=1)}{P(X|C=0)} &= \log\int_l P(l_T=l|C=1) P(X_l|C_l=1) \int_\phi \frac{P(\phi|X)}{P(X_l|\phi,C_l=0)} \\
 &= \softmax{l=1,\ldots,M}{\log P(l_T=l|C=1) + \log P(X_l|C_l=1) + \softmax{\phi\in\Phi}{ \log P(\phi|X) - \log P(X_l|\phi,C_l=0)}} \\
 \end{align*}
 assuming uniform prior on target locations $P(l_T=l|C=1)$ and on target type $P(Y_l=\theta|C_l=1)$:
 \begin{align*}
 \log\frac{P(X|C=1)}{P(X|C=0)} &= \softmax{l=1,\ldots,M}{ \softmax{\theta\in\Theta_1}{L_{\Theta_1}(X_l)}  
 + \softmax{\phi\in\Phi}{ -\softmax{\theta\in\Theta_0}{L_\theta(X_l) + \log \phi_\theta} + \log P(\phi|X)} }-\log(M|\Theta_1|)
 \end{align*}
 which proves Equation~\ref{eq:hetero-search},\ref{eq:hetero-search-target} and \ref{eq:hetero-search-distractor}.
 
\subsection{State formulation in visual search} \label{me:state}
We have chosen the log posterior ratios at all locations: $\vec{Z}:Z_l(t)=\log\frac{P(X_l|C_l=1)}{P(X_l|C_l=0)}, l=1\ldots M$, to be the state of our model because the resultant system is Markov: i.e. $\vec{Z}$ is a sufficient statistic to compute both the overall log likelihood ratio $S_{\mbox{homo-search}}$ and likelihood of future observations. 

First, as shown in~\cite{Chen2011,ma2011behavior}
\begin{align*}
S_{\mbox{homo-search}} &= \log \frac{P(C=1|X)}{P(C=0|X)} = \softmax{l=1\ldots M}{Z_l} - \log(M)
\end{align*}

Second, let $\Delta X = X(t+1)-X(t)$ denotes new observations at all locations at time $t+1$, the likelihood of $\Delta X$ is obtained by marginalizing the target location $l_T$. Denote $l_T=0$ the target-absent event:
\begin{align*}
P(l_T=0|X(t)) &= P(C=0|X) = \frac{1}{1+\exp(S_{\mbox{homo-search}})} = \frac{1}{1 + \sum_l \exp(Z_l)/M}\\
P(l_T,l_T>0|X(t)) &= \frac{\exp(Z_{l_T})/M}{1 + \sum_l \exp(Z_l)/M}
\end{align*}
For notational convenience, define $Z_0=\log (M)$, then the equations above simplifies to: 
\begin{align*}
P(l_T|X) = \frac{\exp(Z_{l_T})}{\sum_{l=0}^M \exp(Z_l)}
\end{align*}

The posterior on $l_T$ is sufficient to compute likelihood of $\Delta X$:
\begin{align*}
P(\Delta X | X(t)) &= P(\Delta X, C=0|X(t)) + P(\Delta X, C=1|X(t)) \\
\mbox{where\ \ } P(\Delta X,C=0|X(t)) &= P(\Delta X|C=0)P(C=0|X(t)) = P(l_T=0|X(t))\prod_l P(\Delta X_l|C_l=0) \\
P(\Delta X, C=1|X(t)) &= \sum_{l_T} P(\Delta X|l_T) P(l_T|X(t))\\
&= \sum_{l_T} P(\Delta X_{l_T}|C_{l_T}=1) \prod_{l\neq l_T} P(\Delta X_l|C_l=0) P(l_T|X(t)) 
\end{align*}

\subsection{Translating optimal thresholds for discrimination to asymptotic thresholds for search} \label{me:thresholds}
We discuss how to design thresholds for visual search that asymptotically achieves the best ER vs RT trade-off (as in Conjecture.~\ref{conj:vs} and Equation.~\ref{eq:upper-approx} and~\ref{eq:lower-approx}). This is done by relating the asymptotically optimal visual search thresholds $\{\tau_-^{vs}, \tau_+^{vs}\}$ to two other pairs of thresholds: 
\begin{itemize}
\item $\{\tau_-, \tau_+\}$: the optimal thresholds for discrimination with an {\it even} prior ratio (i.e. $P(C=1)/P(C=0)=1$)
\item $\{\tau'_-, \tau'_+\}$: the optimal thresholds for discrimination with a {\it biased} prior ratio of $1/M$.
\end{itemize}

(I) $\{\tau_-^{vs}, \tau_+^{vs}\} = \{\tau'_-, \tau'_+\}$: the asymptotic search thresholds are identical to the discrimination threshold with a $1/M$ prior ratio. The asymptotic case is where the locations $l\neq l^*$ are absolutely sure that they do not contain any target, i.e. $Z_l(t)\rightarrow -\infty, \forall l\neq l^*$. Asymptotically (i.e. after collecting a significant amount of information) this always happens when the target is absent, and happens with probability $1/M$ when the target is present (when $l^*$ is the target location). Therefore, the asymptotic search problem can be reduced to a visual discrimination problem with a prior ratio of $1/M$. 

(II) $\{\tau'_-,\tau'_+\} + \log(1/M) = \{\tau_-,\tau_+\} $: log prior ratio causes an additive change to the optimal discrimination thresholds. Let $\gamma_+$ and $-\gamma_-$ (note that $\gamma_+, \gamma_->0)$ be the upper and lower threshold for visual discrimination with a prior of $p$ for target-present. Let $\mbox{RT}_{C}$ and $\mbox{ER}_{C}$ be the expected response time and error rate when the stimulus type is $C \in \{0,1\}$. The error rates, assuming the two thresholds are far apart, are given by (see summary in ~\cite{palmer2005effect}): 
\begin{align*}
 \mbox{RT}_{1}(\gamma_+, \gamma_-) &\approx  \mbox{RT}_{1}(\gamma_+)= kC_{\mbox{time}}\gamma_+\\
  \mbox{RT}_{0}(\gamma_+, \gamma_-) &\approx \mbox{RT}_{0}(\gamma_-)=kC_{\mbox{time}}\gamma_- \\
 \mbox{ER}_{1}(\gamma_+, \gamma_-) &\approx  \mbox{ER}_{1}(\gamma_-)= \frac{1}{1+e^{\gamma_-}}\\
  \mbox{ER}_{0}(\gamma_+, \gamma_-) &\approx \mbox{ER}_{0}(\gamma_+)=\frac{1}{1+e^{\gamma_+}} 
 \end{align*}
 where $k$ is an unknown constant that is inversely proportional to the drift-rate.  
The total risk $\mathcal{R}(\gamma_+,\gamma_-)$ is given by:
 \begin{align*}
 	\mathcal{R}(\gamma_+,\gamma_-) &= p\mbox{RT}_{1}(\gamma_+) + (1-p)\mbox{RT}_{0}(\gamma_-) + p\mbox{ER}_{1}(\gamma_-) + (1-p)\mbox{ER}_{0}(\gamma_+)\\
 \end{align*}
At the optimal thresholds $\gamma_+^*$ and $\gamma_-^*$, it must be that the local derivatives of the risk function w.r.t. the thresholds are zero:
\begin{align*}
 \frac{\partial \mathcal{R}}{\partial \gamma_+} \left \vert_{\gamma_+=\gamma_+^*} \right. = 0 & \Longrightarrow  kC_{\mbox{time}}= \frac{(1-p)e^{-\gamma_+^*}}{p(1+e^{-\gamma_+^*})^2}  \approx \frac{1-p}{p} e^{-\gamma_+^*} = e^{-\left(\gamma_+^* + \log\frac{p}{1-p}\right)}\\
& \Longrightarrow \gamma_+^*(p)= -\log(kC_{\mbox{time}}) - \log \frac{p}{1-p}\\
 \frac{\partial \mathcal{R}}{\partial \gamma_-} \left \vert_{\gamma_-=\gamma_-^*} \right. = 0  &\Longrightarrow  \gamma_-^*(p)= -\log(kC_{\mbox{time}}) + \log \frac{p}{1-p}
\end{align*}

Setting $p=1/2$ (or equivalently, $\log\frac{p}{1-p}=0$) and $p=1/(1+M)$ (or equivalently, $\log\frac{p}{1-p}=-\log(M)$) respectively, we have:
\begin{align*}
	\tau_+ &=  \gamma^*_+(\frac{1}{2}) = -\log(kC_{\mbox{time}}) \\
	\tau'_+ &=  \gamma^*_+(\frac{1}{1+M}) = -\log(kC_{\mbox{time}}) + \log(M) \\
	\Longrightarrow & \tau_+' = \tau_+ + \log(M) \\
	\mbox{Similarly,} &\\
	\Longrightarrow& \tau_-'  = -\gamma^*_-(\frac{1}{1+M})  = -(\gamma^*_-(\frac{1}{2}) - \log(M)) =  \tau_- + \log(M) 
\end{align*}
Therefore, the optimal thresholds $\{\tau'_-,\tau'_+\}$ with a biased prior ratio may be obtained by offsetting the optimal thresholds $\{\tau_-,\tau_+\}$ with the log prior ratio. 

Combining (I) and (II), see see that the asymptotic visual search thresholds are given by $\{\tau_-^{vs}, \tau_+^{vs}\} = \{\tau_-,\tau_+\} + \log(M)$.

\subsection{Dynamic programming} \label{me:dp}
We use dynamic programming (Eq.~\ref{eq:dp}) to solve for the optimal decision strategy for each $(\vec{Z}, t)$ pair. We set up the recurrence relationship in Equation.~\ref{eq:dp} by computing for each $\vec{Z}$:
\begin{align}
	P_1(\vec{Z}) &\defas P(C=0|\vec{Z}) =   \frac{1}{1 + \sum_{l=1}^M \exp(Z_l)/M} \label{eq:p1}\\
	P(\vec{Z}(t+1)|\vec{Z}(t)) &=  \sum_{l_T=0}^M P(\vec{Z}(t+1)|l_T,\vec{Z}(t)) P(l_T|\vec{Z}(t)) \nn \\
	&= \sum_{l_T=0}^M \mathcal{N}(\vec{Z}(t+1)|  \vec{Z}(t) + \Delta \vec{Z}^{(l_T)}, V^{(l_T)}) \frac{\exp(Z_{l_T})}{\sum_{l=0}^M\exp(Z_l)} 
\end{align}
where $\Delta \vec{Z}^{(l_T)}$ and $V^{(l_T)}$ are the mean and variance of the change in $\vec{Z}$ when the target location is $l_T$ (~Eq.~\ref{eq:llr-gaussian-input}). Let $\mu_0$ and $\mu_1$ be the diffusion slopes (Eq.~\ref{eq:ll-gaussian-input}) when the stimulus is a distractor and a target, respectively:
\begin{align*}
	\Delta Z_l^{(l_T)} &= \mu_0(\mu_1-\mu_0)\delta t  \hspace{1in} \forall l \neq l_T \\
	\Delta Z_{l_T}^{(l_T)} &= \mu_1(\mu_1-\mu_0)\delta t \\
	V^{(l_T)}_l &= (\mu_{\Theta_1}-\mu_{\Theta_0})^2 \delta t \hspace{1in} \forall l=1,\ldots,M
\end{align*}

Having readily computed $P_1(\vec{Z})$ and $P(\vec{Z}(t+1)|\vec{Z}(t))$, we use backward induction to solve the dynamic programming equation (Eq.~\ref{eq:dp}) by starting from an infinite horizon $t=\infty$ (see below for practical details), at which point the ideal observer is forced to declare either target-present or target-absent. 

In our experiments we can only handle two display locations ($\vec{Z}=\{Z_1,Z_2\}$). We discretize each $Z_l$ from $-10$ to $10$ into $2000$ bins and time into bins of $10\ ms$. These parameter values are found by searching in log scale the tightest range and the most economic discretization level that do not affect the result. To simulate the infinite horizon we start backward induction from $t=T_{max}$, and increase $T_{max}$ on a log scale until the solution for the first $10$ seconds stabilizes. We find that the optimal decision strategy is constant in time, which is not surprising since the transition probabilities $P(\vec{Z}(t+1)|\vec{Z})$ and the terminal values $P_1(\vec{Z})$ are constant in time. Thus we can stop the induction whenever the decision boundaries converge in time. Typically $T_{max}=15$ seconds is sufficient to handle diffusion slopes  as low as $\pm5/sec$.




\end{document}